2014

# ENTERPRISE SYSTEMS LIFECYCLE-WIDE INNOVATION READINESS

Sachithra Lokuge
*Queensland University of Technology*, s.lokuge@qut.edu.au

Darshana Sedera
*Queensland University of Technology*, d.sedera@qut.edu.au

# ENTERPRISE SYSTEMS LIFECYCLE-WIDE INNOVATION READINESS

Sachithra Lokuge, Science and Engineering Faculty, Queensland University of Technology, Brisbane, Australia, s.lokuge@qut.edu.au

Darshana Sedera, Science and Engineering Faculty, Queensland University of Technology, Brisbane, Australia, d.sedera@qut.edu.au

## Abstract

*Enterprise Systems have been touted as a key driver of delivering benefits through innovation in corporate Information Systems. The advent of such systems expects to deliver best practices that improve organizational performance. Yet, most Enterprise System installations struggle to see lifecycle-wide value of it. Considering that Enterprise Systems deliver lifecycle-wide innovation; we observe organizational readiness for lifecycle-wide Enterprise Systems innovation. The A VICTORY a-priori model compares contributions of eight constructs for organizational readiness for continuous Enterprise Systems innovation. The model is tested responses of both client and implementation partner. Results indicate that six of the eight constructs of readiness make significant contributions to organizational readiness for Enterprise Systems innovation.*

*Keywords: Enterprise Systems, Innovation, Antecedents of Innovation, Innovation Readiness.*

# 1 INTRODUCTION

Globalization has changed the socio-economical background of industries and has transformed the landscape of the modern business (Gorodnichenko et al. 2010), where the changing needs of the customers, lifestyles, together with the advancements in technology, have made innovation a necessity for all businesses (Baregheh et al. 2009).

Many organizations embrace Enterprise Systems (ES) to increase the organizational performance, efficiency and most importantly to attain innovation (Gable et al. 2008; Sedera and Gable 2010). ES purportedly consists of best practices that allow organizations to change their current business processes and organizational structures radically (Seddon et al. 2010). Therefore, the adoption of an ES is considered as a radical change (Kraemmerand et al. 2003). On the other hand, ES investments are under increasing scrutiny and pressure to justify their value, especially considering the substantial resource investment. The post go-live stage is highly critical for organizations to realize returns on ES investments (Bhattacherjee 2001; Jasperson et al. 2005). However, it has been argued that in the post go-live stage, organizations rarely consider ES as an innovation, thus preventing them to realize optimal benefits (Li et al. 2013). The Standish Group reports that fewer than 10% of large ES installations succeed in using the full potential of their systems (Momoh et al. 2010). On the other hand, there have been reports too on organizations achieving high levels of lifecycle-wide success through ES, mostly by focusing it as an enabler of continuous innovation (Seddon et al. 2010).

Thus, it is not surprising that all ES using organizations strive for ES-led innovation. Herein, we argue that innovation is not simply limited to the adoption of an ES; rather ES should enable continuous innovations throughout its lifecycle. However, lifecycle-wide innovation through ES does not happen automatically with the advent of ES. It is something that requires careful planning at the outset and thorough management thereafter. Yet, lifecycle-wide innovation is seldom observed in academic literature. Some scholars suggest that to attain innovation, the ES critical success factors such as top management support, resources availability and knowledge management should be available throughout the ES lifecycle (Sedera and Gable 2010). On the other hand, implementation partners (i.e. consultant and vendor) face immense pressure, as Weeks and Feeny (2008) noted that client organizations are now expecting innovation beyond the introduction of an ES. Though, the advent of an ES is considered as one of the most significant IT innovations (Davenport 1998), rarely do these organizations plan for the lifecycle-wide requirements to drive innovation beyond the go-live phase. The potential for ES innovation, left unattended, will diminish over the time, until its next major upgrade.

In this research-in-progress paper, we argue that lack of lifecycle-wide innovation is due to the lack of organizational readiness at the time of ES implementation. We present our arguments in the following manner: first, we provide a conceptual foundation for innovation in ES lifecycle. Next, we propose the ‘A VICTORY’ model through which we conceive innovation readiness in an organization, presenting the a-priori model. The data analysis provides insights into how the antecedents of innovation measure innovation readiness and potential differences between client and consultants of their views on innovation readiness.

# 2 ES LIFECYCLE AND POTENTIAL FOR INNOVATION

Markus and Tanis (2000) identified three phases in ES lifecycle: (i) implementation, (ii) shakedown and (iii) onwards/upwards. The adoption of the ES transformed the existing business processes, improved them to create business value and thereby introduced new behaviours to the organizational subsystems and its members (Karimi et al. 2007). Thus, this revolutionary process which caused deeper changes in the organizational climate is considered as a radical innovation in the innovation literature (Damanpour 1988). Green et al. (1995) stated that technological uncertainty, technical inexperience, business inexperience and technology cost are four dimensions that can be used to measure the extent of radicalness. Thus, considering these dimensions implementing an ES can be considered as a radical innovation from the adopting organization's point of view. According to Ross and Vitale (2000), after the implementation and after each major upgrade there is a dip in the performance of the organization. During this shakedown phase all the ES users learn the new system and Sumner (2000) identifies the reason for this dip in performance is the lack of expertise of using the system. Some of the challenges of this phase are; inexpertise, sudden changes in job roles, lack of user training, and software related issues (Nah et al. 2001; Niu et al. 2011). After some time the ES users get familiar with the system and attain a level of expertise. This phase is known as the onwards/upwards phase and it is a period where the organization as well as the users of the system are more stable and have reached a maturity level. In an ideal situation, the expert users will suggest new improvements and the organization will continuously improve the ES to compete with the changing environment. However, in the real business world this ideal scenario seldom occurs. The possible reason is that some organizations could believe that once the ES is in-place, there is no need (or difficult to) to make changes to it. Yet, every innovation similar to ES goes through a lifecycle. Every innovation deteriorates as the technology and the market advances and continuous improvements are required to survive in the dynamic environment. Figure 1 depicts the innovations throughout ES lifecycle.

The thick line denotes the extent of innovation throughout the ES lifecycle. As discussed earlier, when an ES is introduced it is considered as a radical change. As Damanpour (1988) stated it requires large amount of new knowledge to do a radical innovation. During the implementation, new modules are added to streamline the business processes. Organizations believe that this ultimately would lead to innovation. The challenge with ES is that even though it comprises of best practices and activities that lead to productivity gains, the level of impact of these standards dwindle with the rapidly changing technology and the customer needs. Therefore, it is important to innovate continuously throughout the ES lifecycle. The current thinking of the organizations is that ES would act as a magic wand to resolve problems related to business performance. It is true and achievable if the organizations envisage the possibility of innovation beyond the implementation time. As depicted in the diagram, throughout the ES lifecycle the innovation degrades until the next upgrade of the ES occurs. The dotted line denotes the possibility of maintaining innovative behaviour throughout the ES lifecycle. During the onwards/upwards phase the users have become experts using the system and they can suggest new improvements. These improvements can be identified as incremental innovations and this type of innovation does not require intense knowledge (Popadiuk and Choo 2006). Yet, organisations do not foresee any type of innovation beyond the ES implementation.

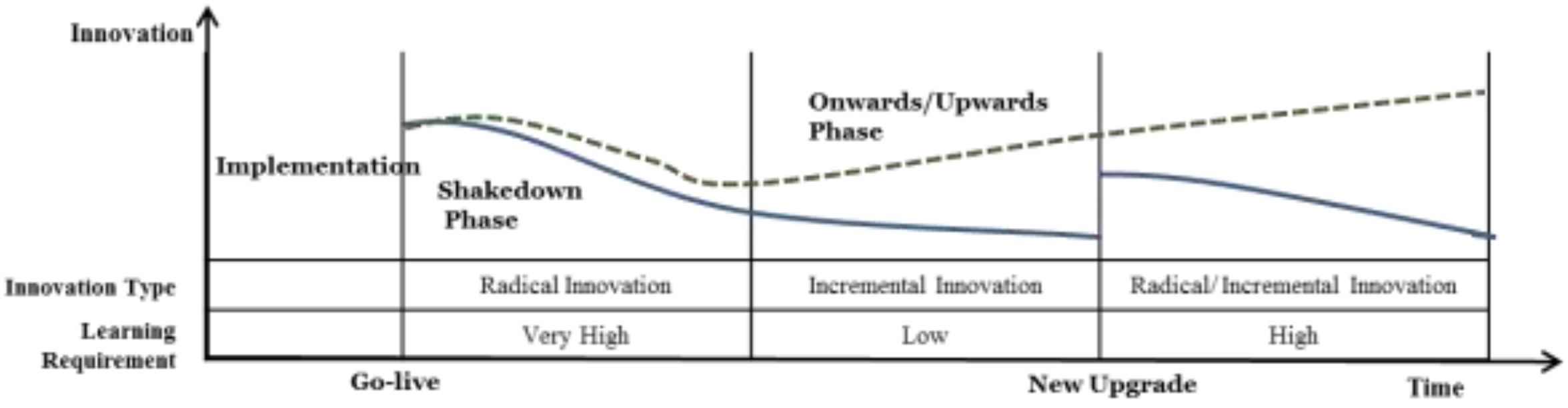

*Figure 1. Innovations in ES lifecycle*

# 3 THE A VICTORY APPROACH TO ASSESS INNOVATION READINESS

We believe the reason for organizations not expecting further innovations after ES implementation is the lack of innovation readiness of these organizations. To measure the innovation readiness we propose the A VICTORY lead approach.

The 'A VICTORY' model is an eight-factor model envisaged as useful for considering organizational behaviour related to the use of new knowledge or innovative practices. This model generally has been used in healthcare environment, yet, Barabba and Zaltman (1991) have applied this model in business studies to seek ways to introduce innovations to organizations (Backer 1995). The A VICTORY model is one of the well-known theories of knowledge utilization and planned change (Johnson 1989). According to Davis (1978) followings are the definitions of each factor.

*Ability* is the resource availability of an organization. Not only the financial resources but also the required human resources, training requirements, and the authority the employees need to carry out change, are considered as determinants of innovation readiness. A key challenge for the organization adopting an ES is the re-deployment and re-skilling of their current employees to the new positions and skills necessary for the new system (Sedera and Dey 2013). Furthermore, training, support staff of the helpdesk, technical upgrades requires adequate fund allocations. Another resource consideration for ES lifecycle is the preparedness for the inevitable upgrades and availability of resources for technical optimizations (Ng et al. 2002).

*Value* observes the characteristics of the organization. Some of the attributes comes under value are open communication, organization culture, and administrative policies. Advent of an ES makes changes to organizational administrative policies and communicational channels. In most cases, communication channels become more formalized through ES workflow and user communications and actions become transparent (Shang and Seddon 2002). However, such policy changes in an organization are not common or continuous; those changes will also be inevitable.

*Information* can be identified as the knowledge or the idea, communication of plan of actions, and clear understanding of the goals of the organization. Davis (1978) stated that "Poor understanding of the details of the change and unsureness of what is expected has been found to be prominent but often unrecognized factors in the failure of change." (p.657). Knowledge management (creating, retention, transfer and application) has been recognized as critical success factor for ES lifecycle-wide success (Sedera and Gable 2010), where they identified that ES stimulate knowledge management as a continuous activity and that needs to be supported by organizational strategy throughout the lifecycle.

*Circumstances* and *Timing* refers to the preparedness towards the changes. An example can be the growing competition in the market, and the advent of new technologies, which decreed the organizations to innovate. Especially, with the current turbulent market and economic conditions warrant organizations to have high degree of agility (Arteta and Giachetti 2004).

*Obligation* can be explained as the motivation or the felt need that makes change acceptable. Sedera and Dey (2013) highlight that ES fail in post go-live due to lack of user motivation to optimize system use. Similar arguments on the importance of motivation for innovative use have been portrayed in Burton-Jones and Straub Jr (2006), Burton-Jones and Gallivan (2007) and Li et al. (2013). They collectively argued that motivation for innovative use of a system would lead to better results. Yet, organizations considering a formal plan for continuous innovation are a minority.

*Resistance* is the associated fear or the aversion towards the change (e.g. fear of economic loss, fear of personal security). Past critical success factor and implementation studies have clearly identified that 'resistance' to change as a key barrier for ES success (Robey et al. 2002). Most studies, following change management models, assumed that resistance diminishes over a period of time (Waddell and Sohal 1998). Yet, the psychological literature demonstrates that unattended resistance would lead to expression of their frustration through other means (Sheth and Stellner 1979). This has been discussed in recent literature (Sedera and Dey 2013; Sedera and Gable 2010), where users create 'boot-leg'/'unauthorized' systems, without using their ES (Sedera and Dey 2007).

*Yield* is the rewarding mechanisms such as incentives, new titles, and group recognition that organizations use to encourage innovation or change. Studies have demonstrated the value of incentive schemes for implementation teams to encourage on-time, on-budget ES implementations. Yet, despite it being valued as an important aspect for the health of ES lifecycle, we are unable to find any studies that reward for innovative behaviours. Eden et al. (2012) identified that, in general, studies reporting post implementation reward is minimal.

# 4 THE A-PRIORI MODEL

By applying the 'A VICTORY' model below, we attempt to measure the innovation readiness of the ES project lifecycle.

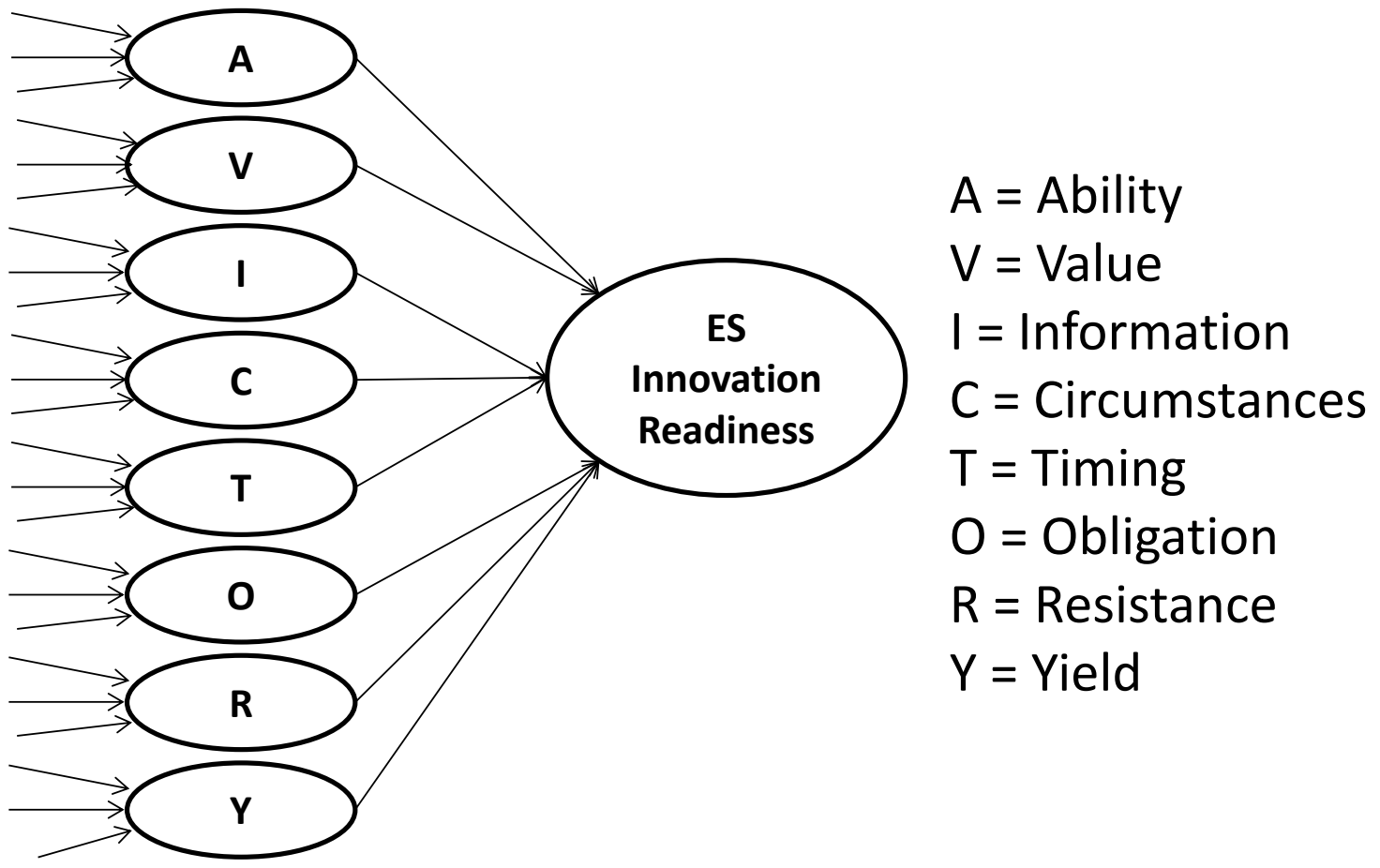

*Figure 2: The a-priori model*

The a-priori model of ES innovation readiness model has eight antecedents– individually they are all conceived and measured as formative. A good formative index– one that exhausts the entire domain of the construct completely, means that the constructs should collectively represent all the relevant aspects of the variable of interest (Bagozzi and Phillips 1982; Fornell and Bookstein 1982). Therefore, its purpose, akin to the function phase of the Burton-Jones and Straub Jr (2006) approach, is to justify the a-priori salient measurement domains as per figure below (i.e. the constructs of A VICTORY) and identify appropriate measures for each dimension of readiness. The a-priori model antecedents; (i) need not co-vary, (ii) are not interchangeable, (iii) cause the core-construct as opposed to being caused by it, and (iv) may have different antecedents and consequences in potentially quite different nomological nets (Cenfetelli and Bassellier 2009; Jarvis et al. 2003; Petter et al. 2007). Moreover, use of formative constructs in this case provide a 'specific and actionable attributes' of a concept (Mathieson et al. 2001), which is particularly interesting from a practical viewpoint as the weight of the construct can be used to draw practical implications on the importance of specific details and therefore guide practical enforcement on the characteristics (See (Furneaux and Wade 2011)).

# 5 DATA COLLECTION AND ANALYSIS

Our preliminary data was gathered from 82 members of the ES implementation team. It included 40 members from the implementation partner and 42 from the client organization. The organization (henceforth referred to as SCM-company to protect anonymity covered by the university ethics agreement) decided to implement SAP Financials and Controlling (SAP-FI/CO), Materials

Management (SAP-MM), Human Capital Management (SAP-HCM) and Supply Chain Management (SAP-SCM) modules in late 2013. The objectives of adoption, time phase and the scope of the implementation is consistent with recent market surveys (Kimberling 2013). In general, similar to most organizations, SCM-company thought of ES as a long-term strategic investment. Table 1 reports descriptive statistics of SCM-company. Most respondents from the client organization represent the management or senior management level.

| | *Details* | | *Details* |
|---|---|---|---|
| **Revenue in 2012-2013** | *USD $52.6 Million* | **New modules considered** | *Sales and Distribution (2015)* |
| **Industry Sector** | *Manufacturing* | **Implementation Approach** | *Phased* |
| **Expected number of users** | *71 (Phase 1) 120 (Phase 2)* | **Level of customization** | *Medium* |
| **Expected completion (months)** | *7 months (Phase 1)* | **Size of implementation team** | *91 (50 client; 41consultant)* |

*Table 1: Details of the ES implementation*

The survey items were derived through the literature review, summarized in Appendix A. The table in Appendix A consists of 28 antecedents of innovation derived through 50 studies (due to space limitations we only demonstrate 35 studies). Having derived the constructs from the original studies, we then map each construct to one of the eight constructs of the A VICTORY model. Two researchers conducted the mapping exercise, eventually arriving at 100% agreement. The review of innovation literature revealed that innovation in general is influenced by environmental, organizational and individual factors (Damanpour 1991; Kimberly and Evanisko 1981).

Using guidelines of Cook et al. (1979) and Diamantopoulos (2010) the pool of items further strengthens the derivation of an instrument for formative constructs. We employed IBM SPSS version 25 and SmartPLS 2.0 (Ringle et al. 2005) in our data analysis. Partial least squares tests (Wold 1989) is a structural equation modelling (SEM) technique that is well suited for highly complex predictive models, supports the mapping of formative observed variables and smaller sample sizes (Becker et al. 2012; Chin et al. 1988; Gefen et al. 2000; Henseler and Sarstedt 2013).

*Common Method Bias:* Sharma et al. (2009) argue against the common practice of gathering perceptual data on both the independent variable and the dependent variable from the same respondent, as it may create Common Method Variance (CMV)[1]. However, as observed in Gorla et al. (2010), CMV is more likely to exist in abstract constructs (e.g. attitude), compared to concrete measures associated with innovation. Yet, in attention to reducing CMV, items for readiness and its antecedents were not grouped under their construct headings in the survey. We also employed the Herman's one-factor-test resulting that, not all measures leading to a single factor solution – confirming that CMV is unlikely.

*Construct Validity:* Following the guidelines of Cenfetelli and Bassellier (2009), Diamantopoulos and Siguaw (2006) and Diamantopoulos and Winklhofer (2001), we first test for multi-collinearity amongst the measures using Variance Inflation Factors (VIF). The VIF from a regression of all constructs ranged between 1.1 and 2.2, indicating that no significant multi-collinearity exists.

*Testing the Structural Model:* The test of the structural model includes, estimates of the path coefficient, which indicate the strength of the relationship between the independent and dependent variable. And also, the $R^2$ values, which represent the amount of variance explained by the independent variable/s. Together, the $R^2$ and the path coefficient (loadings and significance) indicate how well the data supports the hypothesized model (Wixom and Todd 2005). Figure 3 depicts the structural model with path coefficient ($\beta$) of innovation readiness[2]. The $R^2$ values for the dependent variable were significant at level of 0.005 alpha. Supporting our prepositions, further validating the readiness construct, results show the following:(i) Ability, Value, Circumstances, Timing, Obligation and Yield all are strong-significant predictors of Innovation Readiness; (ii) Our analysis does not support 'Information' and 'Resistance' as strong-significant predictors of Innovation Readiness (grey construct with dotted line). However, the significant independent variables explain at least 65% of the variance of the dependent variable (with $R^2$ s for the dependent variables exceeding 0.65). Table 2 provides results of an independent sample t-test that compare the eight antecedents of the A VICTORY model against the two main parties (client and the consultants) of ES implementation. Results of Table 1 show that client and vendor disagree with Ability, Information, Obligation, Resistance and Yield.

[1] The rationale here is that when gathering data both IV and DV from the same respondent, spurious correlations could result (due to the common method used in data collection), which cannot be necessarily be attributed to the underlying phenomena being tested.

[2] The reliability of the Enterprise Systems Readiness measures was 0.901 (at 0.005 confidence level).

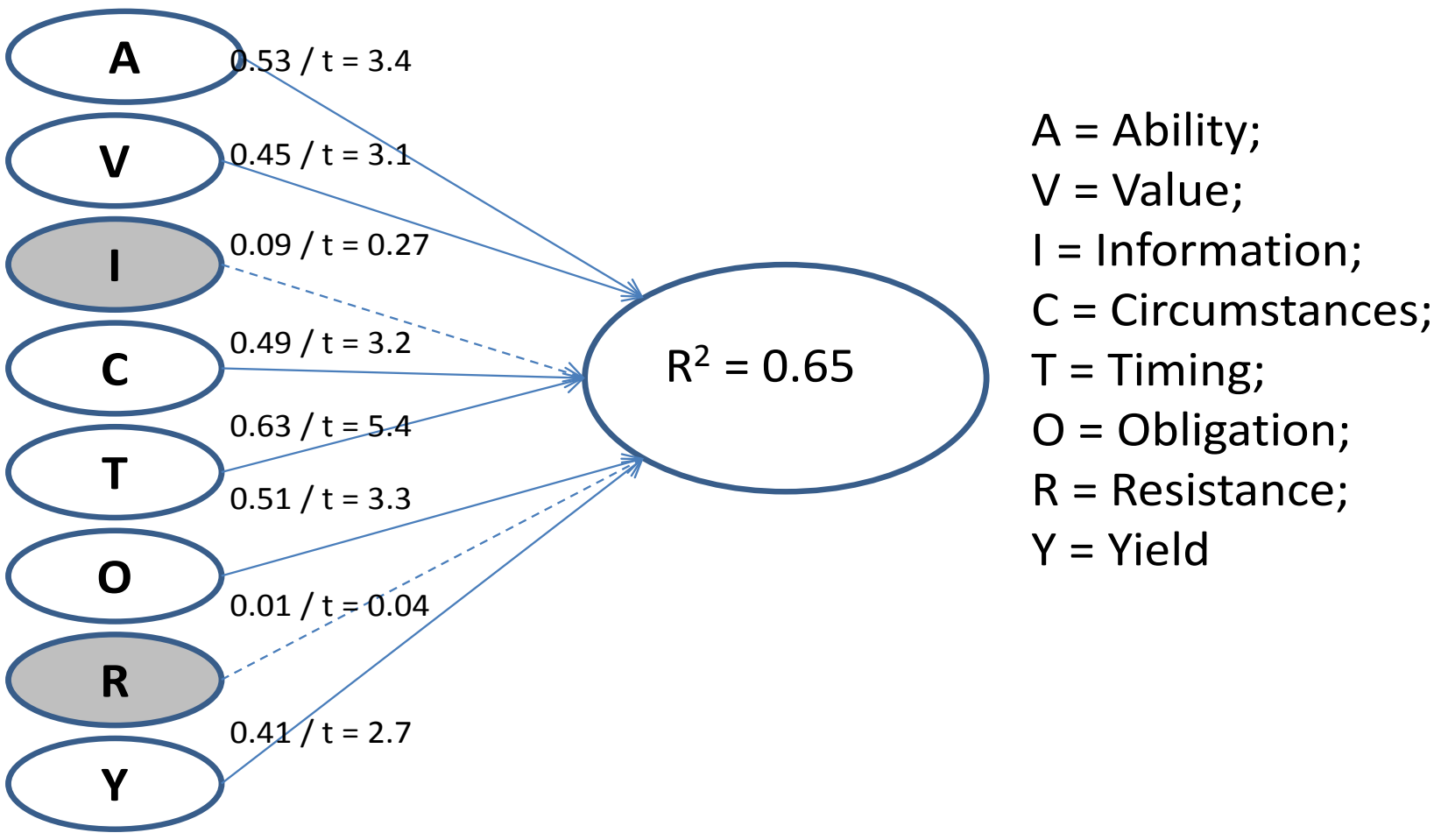

*Figure 3: Results of the PLS analysis*

| | A | V | I | C | T | O | R | Y |
|---|---|---|---|---|---|---|---|---|
| | Sig / t-value* | Sig / t-value* | Sig / t-value* | Sig / t-value* | Sig / t-value* | Sig / t-value* | Sig / t-value* | Sig / t-value* |
| Client Vs. Consultant | 0.02 / -2.41 | 0.71 / 0.86 | 0.01 / -3.35 | 0.65 / .69 | 0.66 / 0.69 | 0.01 / -2.86 | 0.01 / -3.35 | 0.03 / -1.89 |
| * significant at 0.05 | | | | | | | | |

*Table 2: Independent sample t-test results*

# 6 DISCUSSION

The objective of this research-in-progress paper was to test a model for organizational readiness for ES lifecycle-wide innovation. We employed the A VICTORY model of innovation readiness of Davis (1978) to ascertain client readiness for lifecycle-wide innovation through ES. Data was collected from both client and consultant. We found that Information and Resistance do not contribute to ES innovation readiness. The remaining A VICTORY model constructs explained 65% of the variance of organizational innovation readiness. The Information construct is not significant. From the outset, it would seem that "knowledge" is important to the organization for innovation. Yet, as suggested in Figure 1, the organizational learning requirements are less in relation to incremental innovation – thus making the "Information" construct less relevant for innovation readiness. Secondly, the "resistance" construct also shown as non-significant. We argue that 'Resistance' is somewhat contradicting to the notions of innovation. In general, especially with the implementation team as the respondent sample, it is unlikely that resistance is recognized as a barrier for innovation readiness. Though the initial findings are heartening, further research is underway to extend generalizability. Through our conceptualization, the model provides a clear outline of factors important for ES innovation for practitioners, and could facilitate a new track of research on continuous ES innovation. Amongst its limitations, the single site data collection prohibits extensive generalizability.

# 7 APPENDIX A

| Model Factor | Antecedent | (Menzel 1960) | (Greenhalgh et al. 2004) | (Oke et al. 2012) | (Popadiuk et al. 2006) | (Gumusluoğlu et al. 2009) | (Stock et al. 2013) | (Patanakul et al. 2012) | (Robeson et al. 2013) | (Ford et al. 2000) | (Ecker et al. 2013) | (Shane et al. 1995) | (Damanpour et al. 2012) | (Damanpour 1991) | (Kessler et al. 1996) | (Hartmann et al. 2013) | (Damanpour et al. 2006) | (Mate-Sanchez-Val et al. 2014) | (Backmann 2013) | (Lengnick-Hall 1992) | (Evanschitzky et al. 2012) | (Shu et al. 2012) | (Cooper 2011) | (Büschgens et al. 2013) | (Im et al. 2013) | (Taylor et al. 2006) | (Lu 2012) | (Drazin et al. 1996) | (Somech et al. 2013) | (Jansen et al. 2006) | (Troilo et al. 2014) | (Wang et al. 2013) | (Grigoriou et al. 2013) | (Fey et al. 2005) | (Kuester et al. 2013) | (Berends et al. 2013) |
|---|---|---|---|---|---|---|---|---|---|---|---|---|---|---|---|---|---|---|---|---|---|---|---|---|---|---|---|---|---|---|---|---|---|---|---|---|
| | Availability of Resources | | x | x | x | x | x | x | x | | x | | | | | | | x | | | x | | x | | | x | x | x | | | x | x | | x | x | x |
| | Budget | | | | | x | | | | | x | | | | | | | x | | | | | | | | | x | | | | x | x | | | | x |
| | Skilled Employees | | x | | x | x | | x | | | | | | | | | | x | | | | | x | | x | | | | | | | | | | x | x |
| | Functional Differentiation | | x | | x | | x | x | x | | | | | x | x | | | | | x | x | x | | x | | x | | | x | | x | x | | | x | |
| | Professionalism | | | | | | | | | | x | x | | x | | | | | | | | | | | | | | | | | | | | | | |
| Ability | Project Size | | x | x | | | x | x | x | x | | | x | | x | | x | | | x | x | x | | | x | x | x | | | | x | | | x | | x |
| | Formalization | | x | | | | | | | x | x | x | x | x | x | x | | | x | | x | | | | x | | | | | x | | | | | | x |
| | Centralization | | x | | | | | | | | x | x | x | x | | | | | | | x | | | | | | | | | x | | | | | | |
| | Organization Culture | x | x | | | x | x | x | | x | x | x | | | x | x | | | x | x | x | x | x | x | | | x | x | x | | x | x | x | | x | |
| | Organization Structure | | x | | x | x | | x | | x | x | x | | | | | | x | | x | x | x | x | x | | | | x | | | | x | x | | x | x |
| | Organizational Processes | | | x | x | x | x | | | x | | x | | | | | | x | | x | x | | | | x | | | | | | | | | | x | |
| Value | Connectedness | x | | | | | | x | | | | x | | | x | | | | x | | | x | | x | x | | | | x | x | x | x | x | | | |
| | Specialization | | | | | | | | x | | | | | x | x | | | | | | x | | | | | x | | | x | | | | x | x | x | |
| | Communication | x | | | | | | | | | | | | | | | | | x | x | x | x | x | | | | | | | | | | x | | | |
| | Clarity of Goals | | x | x | | | | | | | | | | | x | | | | x | x | | | x | | | | | | x | | | x | | | | x |
| | Knowledge Sharing | | x | | | | | | | | | | | | | | | | | | | | | | | | | | | | | | | | | |
| Information | Technical Knowledge | | x | | x | | x | | | x | | | | x | | x | | | x | x | x | x | | | | | | | x | x | x | x | x | x | x | x |
| | Environmental Dynamism | | | x | | | x | x | x | | | x | | | x | | | x | x | | x | | | | x | | | x | | x | x | | | | | |
| Circumstances | Environmental Competitiveness | | | x | | | x | | x | | | x | x | | x | | | x | x | | x | x | | | x | | | x | | x | | | | | | x |
| | Timing | | | x | x | x | | | x | | x | | | | | | | | | x | x | | | | | | | | | | x | | | | | |
| Timing | New Technology | | | x | | | | x | | | x | | | | | x | | x | x | | x | | | | | | | | | | | | | x | x | |
| | Management Attitude | | x | x | | x | | x | | x | | | | x | x | | x | | | | x | x | x | | | | | x | x | | x | | | | | |
| | Leadership | x | | | | | x | | | | x | | | | | x | | x | | | | | | x | | | x | | | | | | | | x | |
| Obligation | Management Experience | | | | | | | | x | x | | | x | x | x | x | x | | x | | | | | | | x | | | | | | | x | | | |
| | Project Complexity | | | | | | | | | x | x | | x | | x | | x | | | | | | | | | | | | | | | | | | | |
| Resistance | Managers' Education | | | | | | | x | x | x | | | x | | x | | | | | | | | | | | | | | | | | | | | | |
| | Reward System | | | | | x | x | | | | | | | | x | x | | | | | | | x | x | x | | | | | | | x | | | | |
| Yield | HR Practices | | | x | | x | x | | | | | | | | | | | | | | | | | | | | | | | | | | | | | |

# 8 REFERENCES

Arteta, B., and Giachetti, R. 2004. "A Measure of Agility as the Complexity of the Enterprise System," *Robotics and Computer-Integrated Manufacturing* (20:6), pp. 495-503.

Backer, T.E. 1995. "Assessing and Enhancing Readiness for Change: Implications for Technology Transfer." NIDA research monograph, p. 21.

Backmann, J. 2013. "Antecedents of Interorganizational New Product Development Project Performance: A Meta-Analysis," *Academy of Management Proceedings*: Academy of Management, p. 16044.

Bagozzi, R.P., and Phillips, L.W. 1982. "Representing and Testing Organizational Theories: A Holistic Construal," *Administrative Science Quarterly*), pp. 459-489.

Barabba, V., and Zaltman, G. 1991. *Hearing the Voice of the Market*. Cambridge, MA: Harvard Business School Press.

Baregheh, A., Rowley, J., and Sambrook, S. 2009. "Towards a Multidisciplinary Definition of Innovation," *Management Decision* (47:8), pp. 1323-1339.

Becker, J.-M., Klein, K., and Wetzels, M. 2012. "Hierarchical Latent Variable Models in Pls-Sem: Guidelines for Using Reflective-Formative Type Models," *Long Range Planning* (45:5), pp. 359-394.

Berends, H., Jelinek, M., Reymen, I., and Stultiëns, R. 2013. "Product Innovation Processes in Small Firms: Combining Entrepreneurial Effectuation and Managerial Causation," *Journal of Product Innovation Management*).

Bhattacherjee, A. 2001. "Understanding Information Systems Continuance: An Expectation-Confirmation Model," *MIS Quarterly* (25:3), pp. 351-370.

Burton-Jones, A., and Gallivan, M.J. 2007. "Toward a Deeper Understanding of System Usage in Organizations: A Multilevel Perspective," *Mis Quarterly* (31:4), pp. 657-679.

Burton-Jones, A., and Straub Jr, D.W. 2006. "Reconceptualizing System Usage: An Approach and Empirical Test," *Information Systems Research* (17:3), pp. 228-246.

Büschgens, T., Bausch, A., and Balkin, D.B. 2013. "Organizational Culture and Innovation: A Meta-Analytic Review," *Journal of Product Innovation Management*).

Cenfetelli, R.T., and Bassellier, G. 2009. "Interpretation of Formative Measurement in Information Systems Research," *Mis Quarterly* (33:4).

Chin, J.P., Diehl, V.A., and Norman, K.L. 1988. "Development of an Instrument Measuring User Satisfaction of the Human-Computer Interface," *Proceedings of the SIGCHI conference on Human factors in computing systems*: ACM, pp. 213-218.

Cook, T.D., Campbell, D.T., and Day, A. 1979. *Quasi-Experimentation: Design & Analysis Issues for Field Settings*. Houghton Mifflin Boston.

Cooper, R.G. 2011. "Perspective: The Innovation Dilemma: How to Innovate When the Market Is Mature," *Journal of Product Innovation Management* (28:s1), pp. 2-27.

Damanpour, F. 1988. "Innovation Type, Radicalness, and the Adoption Process," *Communication Research* (15:5), pp. 545-567.

Damanpour, F. 1991. "Organizational Innovation: A Meta-Analysis of Effects of Determinants and Moderators," *Academy of Management Journal* (34:3), pp. 555-590.

Damanpour, F., and Aravind, D. 2012. "Managerial Innovation: Conceptions, Processes, and Antecedents," *Management and Organization Review* (8:2), pp. 423-454.

Damanpour, F., and Schneider, M. 2006. "Phases of the Adoption of Innovation in Organizations: Effects of Environment, Organization and Top Managers1," *British Journal of Management* (17:3), pp. 215-236.

Davenport, T.H. 1998. "Putting the Enterprise into the Enterprise System," *Harvard business review* (76:4), p. 121.

Davis, H.A. 1978. "Management of Innovation and Change in Mental Health Services," *Hosp Community Psychiatry* (29:10), pp. 649-658.

Diamantopoulos, A. 2010. "Reflective and Formative Metrics of Relationship Value: Response to Baxter's Commentary Essay," *Journal of Business Research* (63:1), pp. 91-93.
Diamantopoulos, A., and Siguaw, J.A. 2006. "Formative Versus Reflective Indicators in Organizational Measure Development: A Comparison and Empirical Illustration," *British Journal of Management* (17:4), pp. 263-282.
Diamantopoulos, A., and Winklhofer, H.M. 2001. "Index Construction with Formative Indicators: An Alternative to Scale Development," *Journal of Marketing Research* (38:2), pp. 269-277.
Drazin, R., and Schoonhoven, C.B. 1996. "Community, Population, and Organization Effects on Innovation: A Multilevel Perspective," *Academy of Management Journal* (39:5), pp. 1065-1083.
Ecker, B., van Triest, S., and Williams, C. 2013. "Management Control and the Decentralization of R&D," *Journal of Management* (39:4), pp. 906-927.
Eden, R., Sedera, D., and Tan, F. 2012. "Archival Analysis of Enterprise Resource Planning Systems: The Current State and Future Directions," *Proceedings of the 33rd International Conference on Information Systems*, Orlando Florida, USA: AIS Electronic Library (AISeL).
Evanschitzky, H., Eisend, M., Calantone, R.J., and Jiang, Y. 2012. "Success Factors of Product Innovation: An Updated Meta-Analysis," *Journal of Product Innovation Management* (29:S1), pp. 21-37.
Fey, C.F., and Birkinshaw, J. 2005. "External Sources of Knowledge, Governance Mode, and R&D Performance," *Journal of Management* (31:4), pp. 597-621.
Ford, C.M., and Gioia, D.A. 2000. "Factors Influencing Creativity in the Domain of Managerial Decision Making," *Journal of Management* (26:4), pp. 705-732.
Fornell, C., and Bookstein, F.L. 1982. "Two Structural Equation Models: Lisrel and Pls Applied to Consumer Exit-Voice Theory," *Journal of Marketing Research (JMR)* (19:4).
Furneaux, B., and Wade, M. 2011. "An Exploration of Organizational Level Information Systems Discontinuance Intentions," *MIS Quarterly* (35:3).
Gable, G.G., Sedera, D., and Chan, T. 2008. "Re-Conceptualizing Information System Success: The Is-Impact Measurement Model," *Journal of the Association for Information Systems* (9:7).
Gefen, D., Straub, D.W., and Boudreau, M.-C. 2000. "Structural Equation Modeling and Regression: Guidelines for Research Practice," *Communications of the AIS* (4:7), pp. 1-79.
Gorla, N., Somers, T.M., and Wong, B. 2010. "Organizational Impact of System Quality, Information Quality, and Service Quality," *The Journal of Strategic Information Systems* (19:3), pp. 207-228.
Gorodnichenko, Y., Svejnar, J., and Terrell, K. 2010. "Globalization and Innovation in Emerging Markets," *American Economic Journal: Macroeconomics* (2:2), pp. 194-226.
Green, S.G., Gavin, M.B., and Aiman-Smith, L. 1995. "Assessing a Multidimensional Measure of Radical Technological Innovation," *Engineering Management, IEEE Transactions on* (42:3), pp. 203-214.
Greenhalgh, T., Robert, G., Macfarlane, F., Bate, P., and Kyriakidou, O. 2004. "Diffusion of Innovations in Service Organizations: Systematic Review and Recommendations," *Milbank Quarterly* (82:4), pp. 581-629.
Grigoriou, K., and Rothaermel, F.T. 2013. "Structural Microfoundations of Innovation the Role of Relational Stars," *Journal of Management*), p. 0149206313513612.
Gumusluoğlu, L., and Ilsev, A. 2009. "Transformational Leadership and Organizational Innovation: The Roles of Internal and External Support for Innovation*," *Journal of Product Innovation Management* (26:3), pp. 264-277.
Hartmann, M., Oriani, R., and Bateman, H. 2013. "Exploring the Antecedents to Business Model Innovation: An Empirical Analysis of Pension Funds," Working Paper.
Henseler, J., and Sarstedt, M. 2013. "Goodness-of-Fit Indices for Partial Least Squares Path Modeling," *Computational Statistics* (28:2), pp. 565-580.
Im, S., Montoya, M.M., and Workman, J.P. 2013. "Antecedents and Consequences of Creativity in Product Innovation Teams," *Journal of Product Innovation Management* (30:1), pp. 170-185.

Jansen, J.J., Van Den Bosch, F.A., and Volberda, H.W. 2006. "Exploratory Innovation, Exploitative Innovation, and Performance: Effects of Organizational Antecedents and Environmental Moderators," *Management Science* (52:11), pp. 1661-1674.

Jarvis, C.B., MacKenzie, S.B., and Podsakoff, P.M. 2003. "A Critical Review of Construct Indicators and Measurement Model Misspecification in Marketing and Consumer Research," *Journal of Consumer Research* (30:2), pp. 199-218.

Jasperson, J.S., Carter, P.E., and Zmud, R.W. 2005. "A Comprehensive Conceptualization of Post-Adoptive Behaviors Associated with Information Technology Enabled Work Systems," *MIS Quarterly* (29:3), pp. 525-557.

Johnson, K.W. 1989. "Knowledge Utilization and Planned Change: An Empirical Assessment of the a Victory Model," *Knowledge in Society* (2:2), pp. 57-79.

Karimi, J., Somers, T.M., and Bhattacherjee, A. 2007. "The Impact of Erp Implementation on Business Process Outcomes: A Factor-Based Study," *Journal of Management Information Systems* (24:1), pp. 101-134.

Kessler, E.H., and Chakrabarti, A.K. 1996. "Innovation Speed: A Conceptual Model of Context, Antecedents, and Outcomes," *Academy of Management Review* (21:4), pp. 1143-1191.

Kimberling, E. 2013. "Can Erp Systems Be Used to Drive Innovation?", 2014, from http://panorama-consulting.com/can-erp-systems-be-used-to-drive-innovation/

Kimberly, J.R., and Evanisko, M.J. 1981. "Organizational Innovation: The Influence of Individual, Organizational, and Contextual Factors on Hospital Adoption of Technological and Administrative Innovations," *Academy of Management Journal* (24:4), pp. 689-713.

Kraemmerand, P., Møller, C., and Boer, H. 2003. "Erp Implementation: An Integrated Process of Radical Change and Continuous Learning," *Production Planning & Control* (14:4), pp. 338-348.

Kuester, S., Schuhmacher, M.C., Gast, B., and Worgul, A. 2013. "Sectoral Heterogeneity in New Service Development: An Exploratory Study of Service Types and Success Factors," *Journal of Product Innovation Management*).

Lengnick-Hall, C.A. 1992. "Innovation and Competitive Advantage: What We Know and What We Need to Learn," *Journal of Management* (18:2), pp. 399-429.

Li, X., Hsieh, J.P.-A., and Rai, A. 2013. "Motivational Differences across Post-Acceptance Information System Usage Behaviors: An Investigation in the Business Intelligence Systems Context," *Information Systems Research* (24:3), pp. 659-682.

Lu, L.-H. 2012. "Financial Slack, Board Composition and the Explorative and Exploitative Innovation Behavior of Firms," *Academy of Management Proceedings*: Academy of Management, pp. 1-1.

Markus, M.L., and Tanis, C. 2000. "The Enterprise Systems Experience–from Adoption to Success," in *Framing the Domains of It Research: Glimpsing the Future through the Past*. Pinna • ex Educational Resources, Cincinnati:  pp. 173-207.

Mate-Sanchez-Val, M., and Harris, R. 2014. "Differential Empirical Innovation Factors for Spain and the Uk," *Research Policy* (43:2), pp. 451-463.

Mathieson, K., Peacock, E., and Chin, W.W. 2001. "Extending the Technology Acceptance Model: The Influence of Perceived User Resources," *ACM SigMIS Database* (32:3), pp. 86-112.

Menzel, H. 1960. "Innovation, Integration, and Marginality: A Survey of Physicians," *American Sociological Review* (25:5), pp. 704-713.

Momoh, A., Roy, R., and Shehab, E. 2010. "Challenges in Enterprise Resource Planning Implementation: State-of-the-Art," *Business Process Management Journal* (16:4), pp. 537-565.

Nah, F.F.-H., Lau, J.L.S., and Kuang, J. 2001. "Critical Factors for Successful Implementation of Enterprise Systems," *Business Process Management Journal* (7:3), pp. 285-296.

Ng, C.S.P., Gable, G.G., and Chan, T. 2002. "An Erp-Client Benefit-Oriented Maintenance Taxonomy," *Journal of Systems and Software* (64:2), pp. 87-109.

Niu, N., Jin, M., and Cheng, J.-R.C. 2011. "A Case Study of Exploiting Enterprise Resource Planning Requirements," *Enterprise Information Systems* (5:2), pp. 183-206.

Oke, A., Walumbwa, F.O., and Myers, A. 2012. "Innovation Strategy, Human Resource Policy, and Firms' Revenue Growth: The Roles of Environmental Uncertainty and Innovation Performance*," *Decision Sciences* (43:2), pp. 273-302.

Patanakul, P., Chen, J., and Lynn, G.S. 2012. "Autonomous Teams and New Product Development," *Journal of Product Innovation Management* (29:5), pp. 734-750.

Petter, S., Straub, D., and Rai, A. 2007. "Specifying Formative Constructs in Information Systems Research," *Mis Quarterly* (31:4), pp. 623-656.

Popadiuk, S., and Choo, C.W. 2006. "Innovation and Knowledge Creation: How Are These Concepts Related?," *International Journal of Information Management* (26:4), pp. 302-312.

Ringle, C.M., Wende, S., and Will, A. 2005. "Smartpls (Version 2.0 (Beta))." March 28, 2007

Robeson, D., and O'Connor, G.C. 2013. "Boards of Directors, Innovation, and Performance: An Exploration at Multiple Levels," *Journal of Product Innovation Management*).

Robey, D., Ross, J.W., and Boudreau, M.-C. 2002. "Learning to Implement Enterprise Systems: An Exploratory Study of the Dialectics of Change," *Journal of Management Information Systems* (19:1), pp. 17-46.

Ross, J.W., and Vitale, M.R. 2000. "The Erp Revolution: Surviving Vs. Thriving," *Information systems frontiers* (2:2), pp. 233-241.

Seddon, P.B., Calvert, C., and Yang, S. 2010. "A Multi-Project Model of Key Factors Affecting Organizational Benefits from Enterprise Systems," *MIS Quarterly* (34:2), pp. 305-328.

Sedera, D., and Dey, S. 2007. "Everyone Is Different! Exploring the Issues and Problems with Erp Enabled Shared Service Initiatives," *Americas Conference on Information Systems* Key Stone Colarado USA.: AIS.

Sedera, D., and Dey, S. 2013. "User Expertise in Contemporary Information Systems: Conceptualization, Measurement and Application," *Information & Management* (50:8), pp. 621-637.

Sedera, D., and Gable, G.G. 2010. "Knowledge Management Competence for Enterprise System Success," *The Journal of Strategic Information Systems* (19:4), pp. 296-306.

Shane, S., Venkataraman, S., and MacMillan, I. 1995. "Cultural Differences in Innovation Championing Strategies," *Journal of Management* (21:5), pp. 931-952.

Shang, S., and Seddon, P.B. 2002. "Assessing and Managing the Benefits of Enterprise Systems: The Business Manager's Perspective," *Information Systems Journal* (12:4), pp. 271-299.

Sharma, R., Yetton, P., and Crawford, J. 2009. "Estimating the Effect of Common Method Variance: The Method-Method Pair Technique with an Illustration from Tam Research," *MIS Quarterly* (33:3).

Sheth, J.N., and Stellner, W.H. 1979. *Psychology of Innovation Resistance: The Less Developed Concept (Ldc) in Diffusion Research*. College of Commerce and Business Administration, University of Illinois at Urbana-Champaign.

Shu, C., Page, A.L., Gao, S., and Jiang, X. 2012. "Managerial Ties and Firm Innovation: Is Knowledge Creation a Missing Link?," *Journal of Product Innovation Management* (29:1), pp. 125-143.

Somech, A., and Drach-Zahavy, A. 2013. "Translating Team Creativity to Innovation Implementation the Role of Team Composition and Climate for Innovation," *Journal of Management* (39:3), pp. 684-708.

Stock, R.M., Totzauer, F., and Zacharias, N.A. 2013. "A Closer Look at Cross-Functional R&D Cooperation for Innovativeness: Innovation-Oriented Leadership and Human Resource Practices as Driving Forces," *Journal of Product Innovation Management*).

Sumner, M. 2000. "Risk Factors in Enterprise-Wide/Erp Projects," *Journal of information technology* (15:4), pp. 317-327.

Taylor, A., and Greve, H.R. 2006. "Superman or the Fantastic Four? Knowledge Combination and Experience in Innovative Teams," *Academy of Management Journal* (49:4), pp. 723-740.

Troilo, G., De Luca, L.M., and Atuahene-Gima, K. 2014. "More Innovation with Less? A Strategic Contingency View of Slack Resources, Information Search, and Radical Innovation," *Journal of Product Innovation Management* (31:2), pp. 259-277.

Waddell, D., and Sohal, A.S. 1998. "Resistance: A Constructive Tool for Change Management," *Management Decision* (36:8), pp. 543-548.
Wang, H., Choi, J., Wan, G., and Dong, J.Q. 2013. "Slack Resources and the Rent-Generating Potential of Firm-Specific Knowledge," *Journal of Management*).
Weeks, M.R., and Feeny, D. 2008. "Outsourcing: From Cost Management to Innovation and Business Value," *California Management Review* (50:4), pp. 127-146.
Wixom, B.H., and Todd, P.A. 2005. "A Theoretical Integration of User Satisfaction and Technology Acceptance.," *Information Systems Research* (16:1), pp. 85-102.
Wold, H. 1989. *Introduction to the Second Generation of Multivariate Analysis*. New York: Paragon House: H. Wold (ed.).